\input{aipcheck}

\documentclass[
    ,final            
  ]
  {aipproc}

\layoutstyle{8x11double}

\begin{document}

\title{Implication of a large $\theta_{13}$ for the Tokai to Kamioka and Korea setup}


\author{F.Dufour}{
  address={Section de Physique, Universit\'{e} de Gen\`{e}ve, 1205 Gen\`{e}ve, Switzerland}
}

\begin{abstract}
In this paper, I present the implications of the large value of
$\theta_{13}$ on the Tokai to Kamioka and Korea setup (T2KK).  I study
the sensitivity of T2KK when using a 750~kW beam (the design
luminosity of T2K planned to be achieved by 2017) and a potential
1.66~MW upgraded beam. In addition I compare the capability of the
T2KK setup with the T2HK letter of intent.
\end{abstract}

\maketitle


\section{Introduction}

\begin{figure}[tbp]
    \includegraphics[width=3.0in]{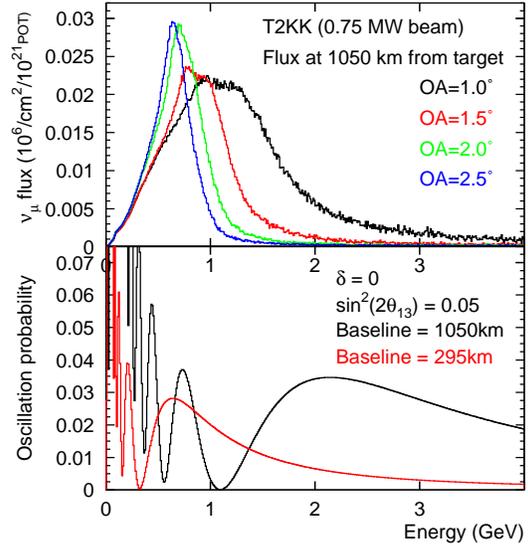}
  \vspace{-1.5pc}
  \caption{(Color online) Neutrino flux as a function of energy for
    several off-axis angles, and a 0.7~5MW beam at 1050~km from the
    target.  For comparison, the $\nu_{\mu}$ $\rightarrow$ $\nu_e$
    probability, for two baselines considered for T2KK (295km and
    1050km). Neutrino mixing parameters are: normal hierarchy, $\Delta
    m^2_{(21,31)}=8.0\times 10^{-5},2.5\times 10^{-3} \mathrm{eV}^2$, and
    $\sin^{2}2\theta_{(12,23)}=0.86,1.0$. We take the earth density to
    be constant and equal to 2.8 ${\rm g/cm^3}$.}
\label{fig:flux_t2kk}
\end{figure}

Previous studies~\cite{Ishitsuka:2005qi,Dufour:2010vr} have shown the
potential of placing two very large Water Cherenkov detectors in the
T2K beam.  By placing one Mton detector at 295~km (in Kamioka) and
another one at about 1000~km (in Korea), one is able to observe the
first and the second oscillation maxima in Kamioka and Korea
respectively.  In addition if the Korean detector is located
approximately on-axis (1 degree is the smallest angle available in
Korea), then the beam is wide-band and the Korean detector will be
able to observe the first and the second oscillation maxima, as it can
be seen in Figure~\ref{fig:flux_t2kk}. Therefore with a detector at
2.5 degree off-axis in Kamioka and a detector at 1 degree off-axis in
Korea, it is possible to measure three types of oscillation maxima
(the first maximum at two baselines, and the second in
Korea)~\cite{Dufour:2010vr}. Such a redundancy is an excellent tool to
measure the mass hierarchy and the CP phase.  Previous studies have
shown this potential, but in this paper, I update the study to account
for the large value of $\theta_{13}$ measured in 2011-2012 and present
its
implications~\cite{Abe:2011sj,Adamson:2011qu,Ahn:2012nd,An:2012eh}.
In addition, I compare my results with those presented in the
Hyper-K letter of intent~\cite{Abe:2011ts}.

\section{Experimental setups}

For this study I assumed two Mton size water Cherenkov detectors. It
was shown previously that a photo-coverage of 20\% is equivalent to a
photo-coverage of 40\%. The capability for rejecting neutral current
background was shown to be around 68\% in both
cases~\cite{Dufour:2010vr}.  In Table~\ref{tbl:expparam} I present the
experimental parameters that were chosen for this study. In addition
to the experimental parameters, I have assumed that
$\sin^{2}2\theta_{(13)}=0.1$ when showing event spectra and I have
assumed that the remaining oscillation parameters are: $\Delta
m^2_{(21,31)}=8.0\times 10^{-5},2.5\times 10^{-3} \mathrm{eV}^2$ and
$\sin^{2}2\theta_{(12,23)}=0.86,1.0$.

\begin{table}[htbp]
\begin{tabular}{|l||c|c|}

\hline
 & T2HK  & T2KK \\
\hline
Volume & 0.56 Mton & 2 $\times$ 0.28 Mton\\
Baseline  &295~km  & 295~km + 1050~km \\
Off-axis angle & $2.5^{\circ}$ &$2.5^{\circ}$+$1.0^{\circ}$ \\
Time ($\nu + \bar{\nu})$&  3+7 years & 5+5 years\\
POT equivalent &  &\\
Beam power & \multicolumn{2}{c|}{0.75 or 1.66~MW }  \\
Proton energy & \multicolumn{2}{c|}{30 GeV }  \\
~~One year &\multicolumn{2}{c|}{$10^7$ seconds} \\
\hline
\end{tabular}
\caption {\label{tbl:expparam}List of experimental parameters}
\end{table}

\section{Potential of a 750~kW beam versus a 1.66~MW beam}

\begin{figure}[ht]
\begin{minipage}[b]{1\linewidth}
\centering
\includegraphics[width=1.1\textwidth]{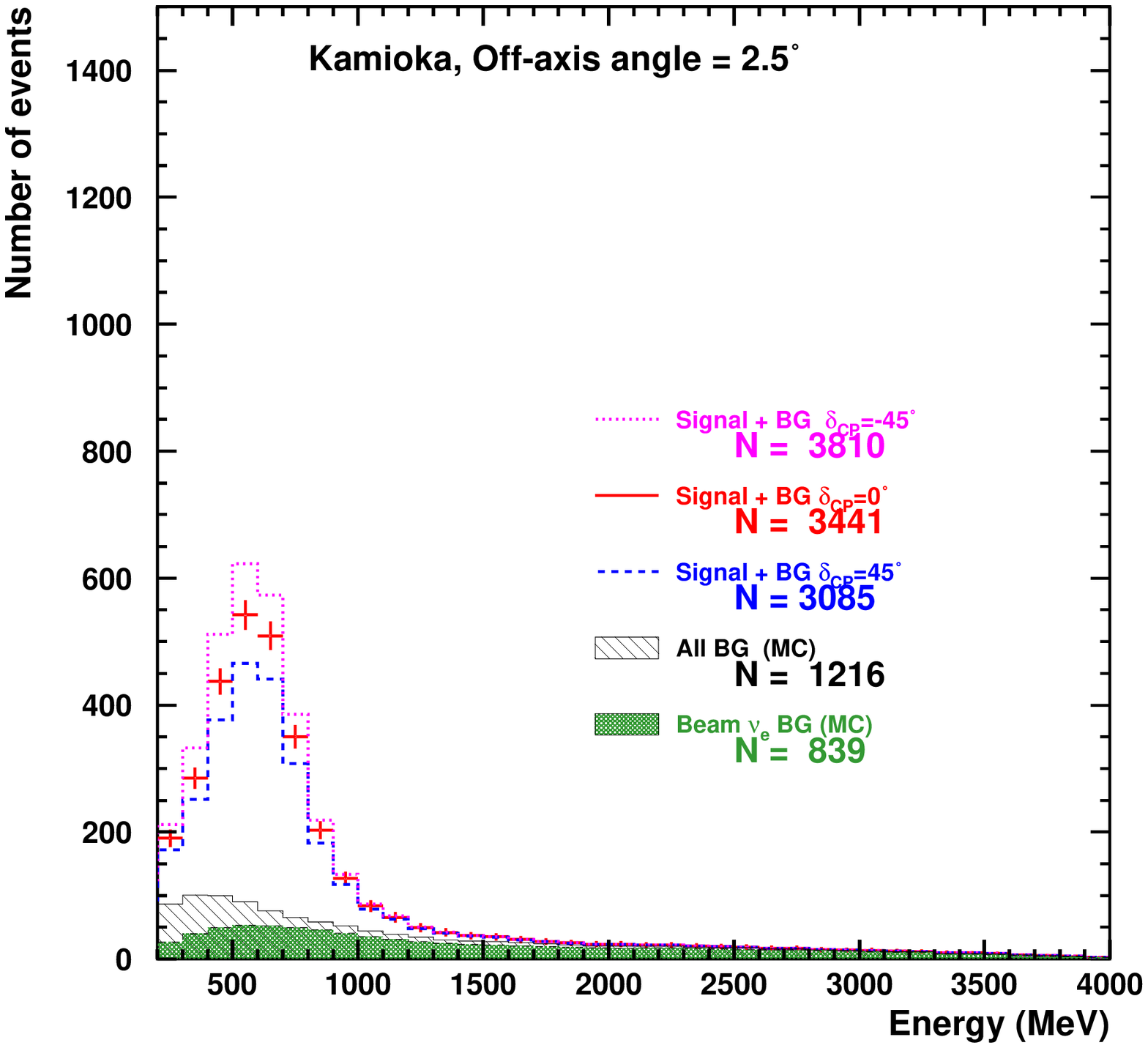}
\includegraphics[width=1.1\textwidth]{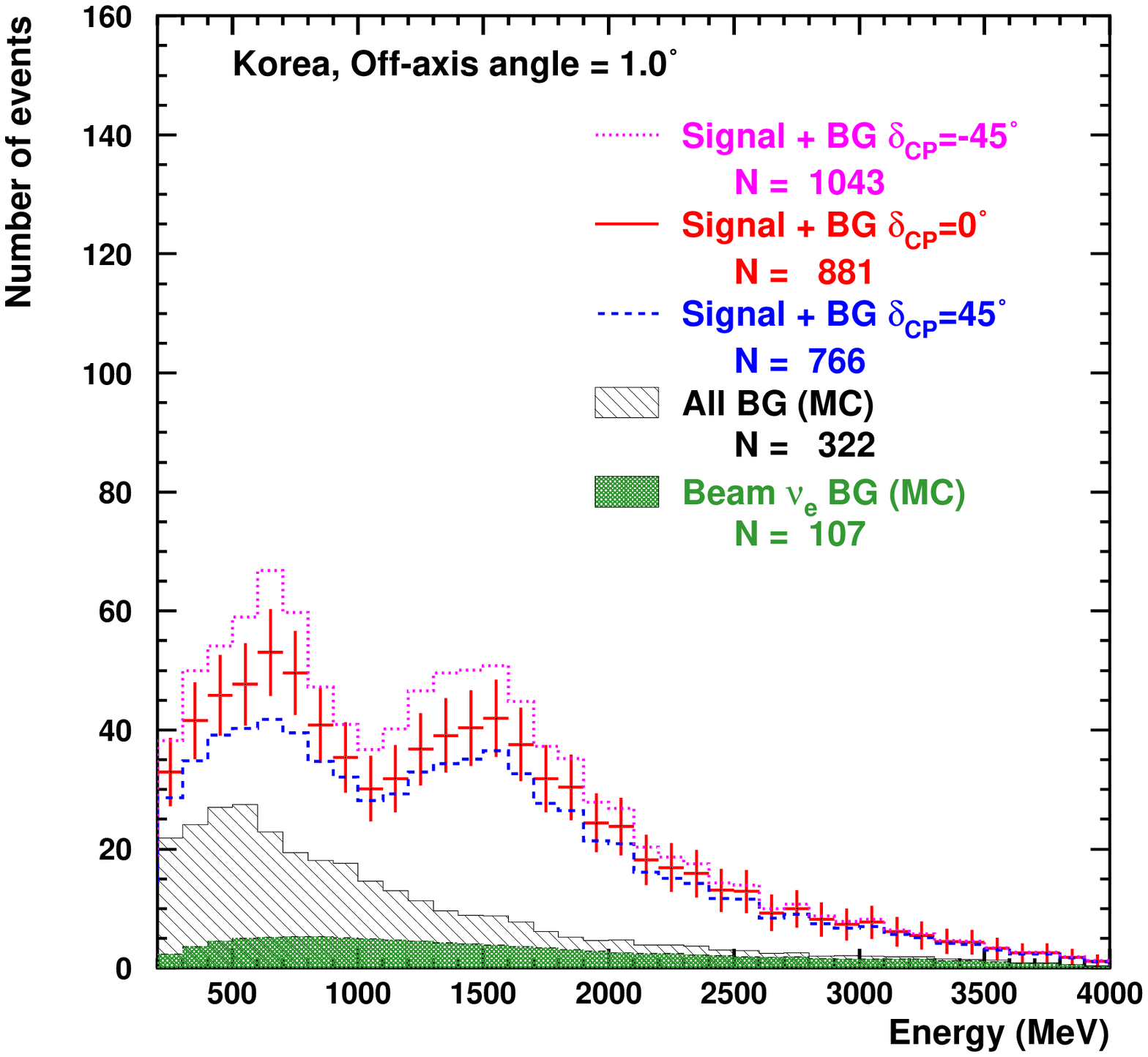}
  \caption{(Color online) Reconstructed energy spectra at Kamioka
    (top) and Korea $1.0^\circ$ off-axis (bottom) for a 750~kW beam
    (left) and a 1.66~MW beam (right) for $\sin^2(2\theta_{13})$ = 0.1
    and normal hierarchy. The remaining oscillation parameters are:
    $\Delta m^2_{(21,31)}=8.0\times 10^{-5},2.5\times 10^{-3} eV^2$
    and $\sin^{2}2\theta_{(12,23)}=0.86,1.0$. Each plot is normalized
    to 5 years of running with neutrino, with 40 GeV protons and in a
    0.27 Mton (FV) detector (i.e. $5 \times 1.17 \times 10^{21}$
    POT).}
\end{minipage}
\hspace{0.5cm}
\begin{minipage}[b]{1\linewidth}
\centering
\includegraphics[width=1.1\textwidth]{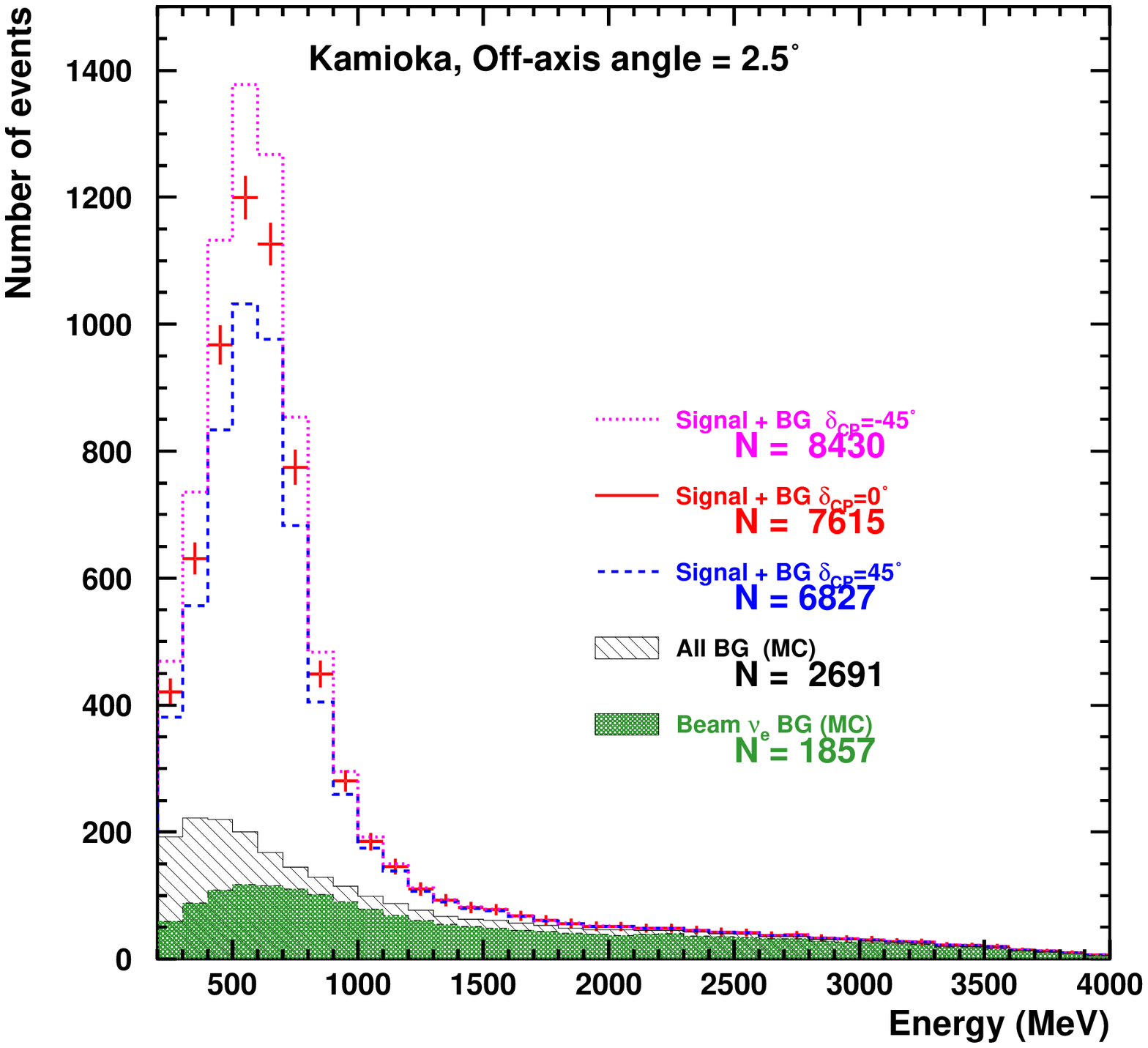}
\includegraphics[width=1.1\textwidth]{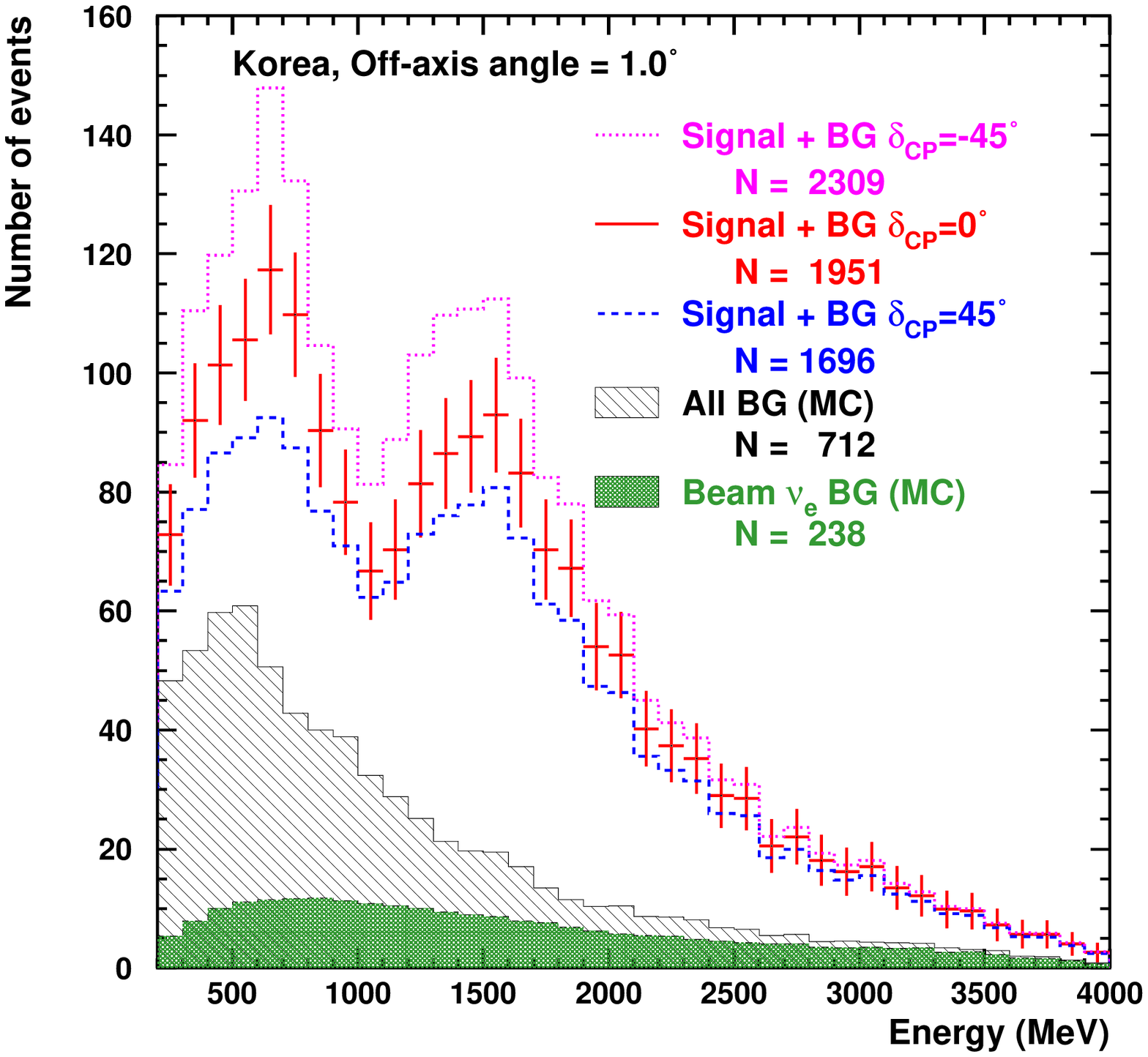}
  \caption{(Color online) Reconstructed energy spectra at Kamioka
    (top) and Korea $1.0^\circ$ off-axis (bottom) for a 750~kW beam
    (left) and a 1.66~MW beam (right) for $\sin^2(2\theta_{13})$ = 0.1
    and normal hierarchy. The remaining oscillation parameters are:
    $\Delta m^2_{(21,31)}=8.0\times 10^{-5},2.5\times 10^{-3} eV^2$
    and $\sin^{2}2\theta_{(12,23)}=0.86,1.0$. Each plot is normalized
    to 5 years of running with neutrino, with 40 GeV protons and in a
    0.27 Mton (FV) detector (i.e. $5 \times 1.17 \times 10^{21}$
    POT).}

\label{fig:spectrum}
\end{minipage}
\end{figure}

In previous studies, the major concern was always the value of
$\theta_{13}$ and therefore we always envisioned that a very high
power beam was necessary. With the large value of $\theta_{13}$
measured in 2012, this is not true anymore and in the case of T2KK,
the 750~kW beam designed for T2K is enough to measure the mass
hierarchy and the CP phase. Of course, if a 1.66~MW beam was built,
the same sensitivity would be achieved twice as fast or a detector
twice as small could be used. Figure~\ref{fig:spectrum} shows the
event spectra in Kamioka (2.5 degree off-axis) and Korea (1 degree
off-axis) for a 750~kW beam and a 1.66~MW beam. The rest of the
parameters are the ones presented in Table~\ref{tbl:expparam}. On
these figures, we can clearly see that the first oscillation maximum
is visible in Kamioka and both the first and second oscillation maxima
are visible at the Korean location.

\section{Comparison between T2KK and T2HK}

\begin{figure*}[htbp]
{\hbox{\hspace{0.0in}
    \includegraphics[width=3.5in]{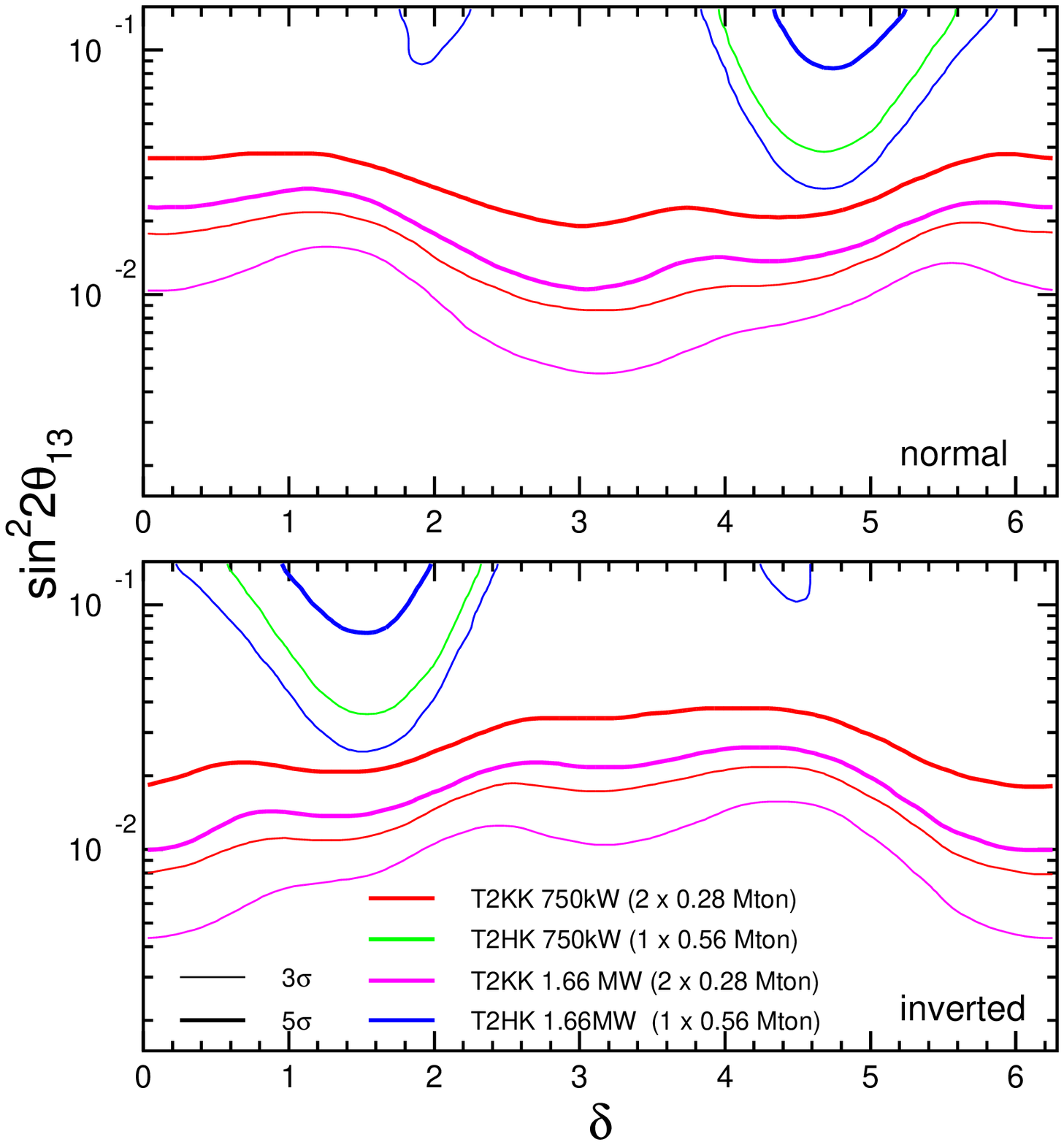}
\hspace{0.0in} \includegraphics[width=3.5in]{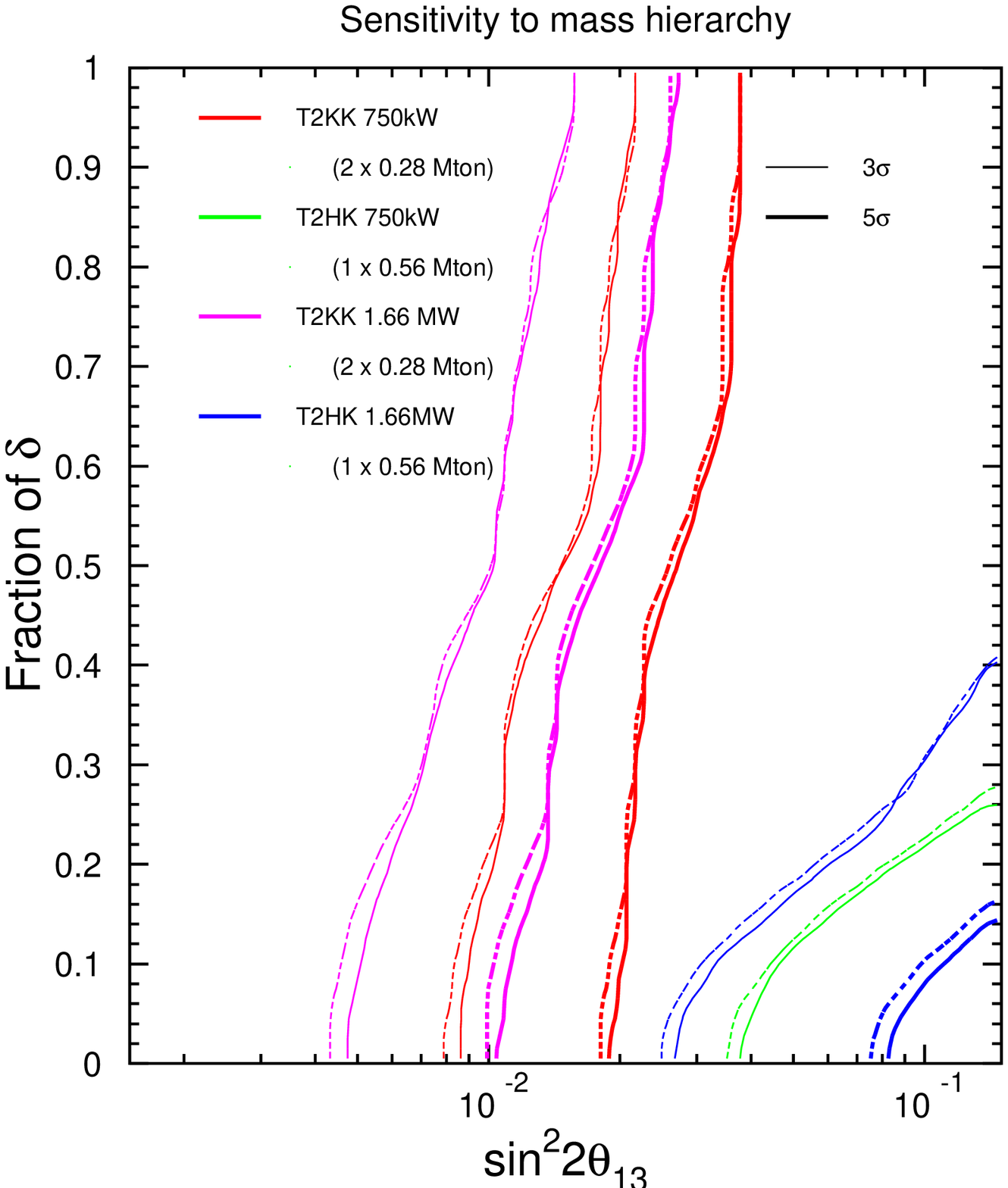}
}} \vspace{-1.5pc} \caption{{\bf Left:} Region of CP phase and
  $\theta_{13}$ where the mass hierarchy can be solved. {\bf Right:}
  Fraction of $\delta_{CP}$ phase space in for which the mass
  hierarchy can be solved. (Plain lines = normal hierarchy, dashed
  lines = inverted hierarchy) }
\label{fig:mass}
\end{figure*}

\begin{figure*}[htbp]
{\hbox{\hspace{0.0in}
    \includegraphics[width=3.5in]{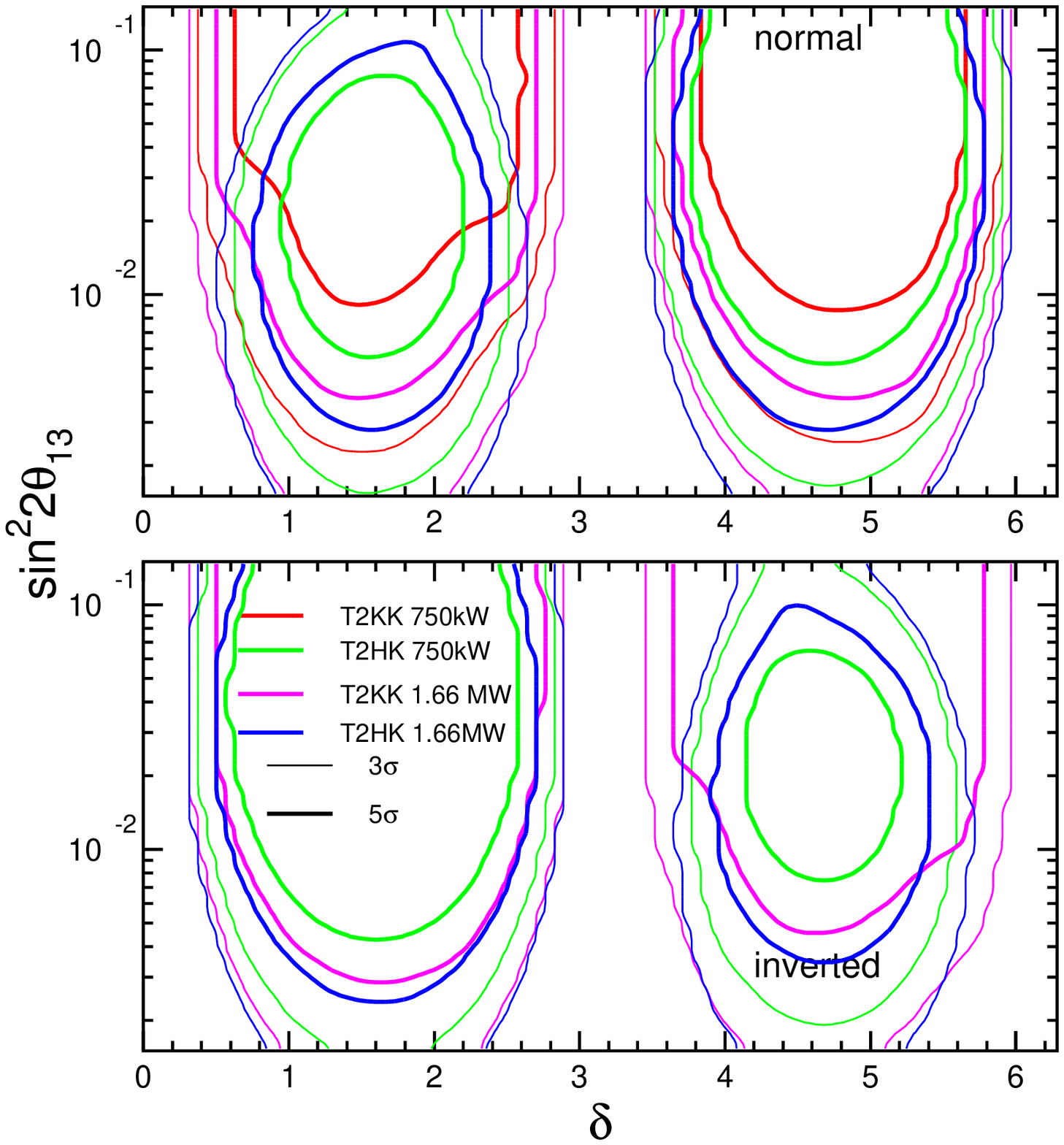}
\hspace{0.0in} \includegraphics[width=3.5in]{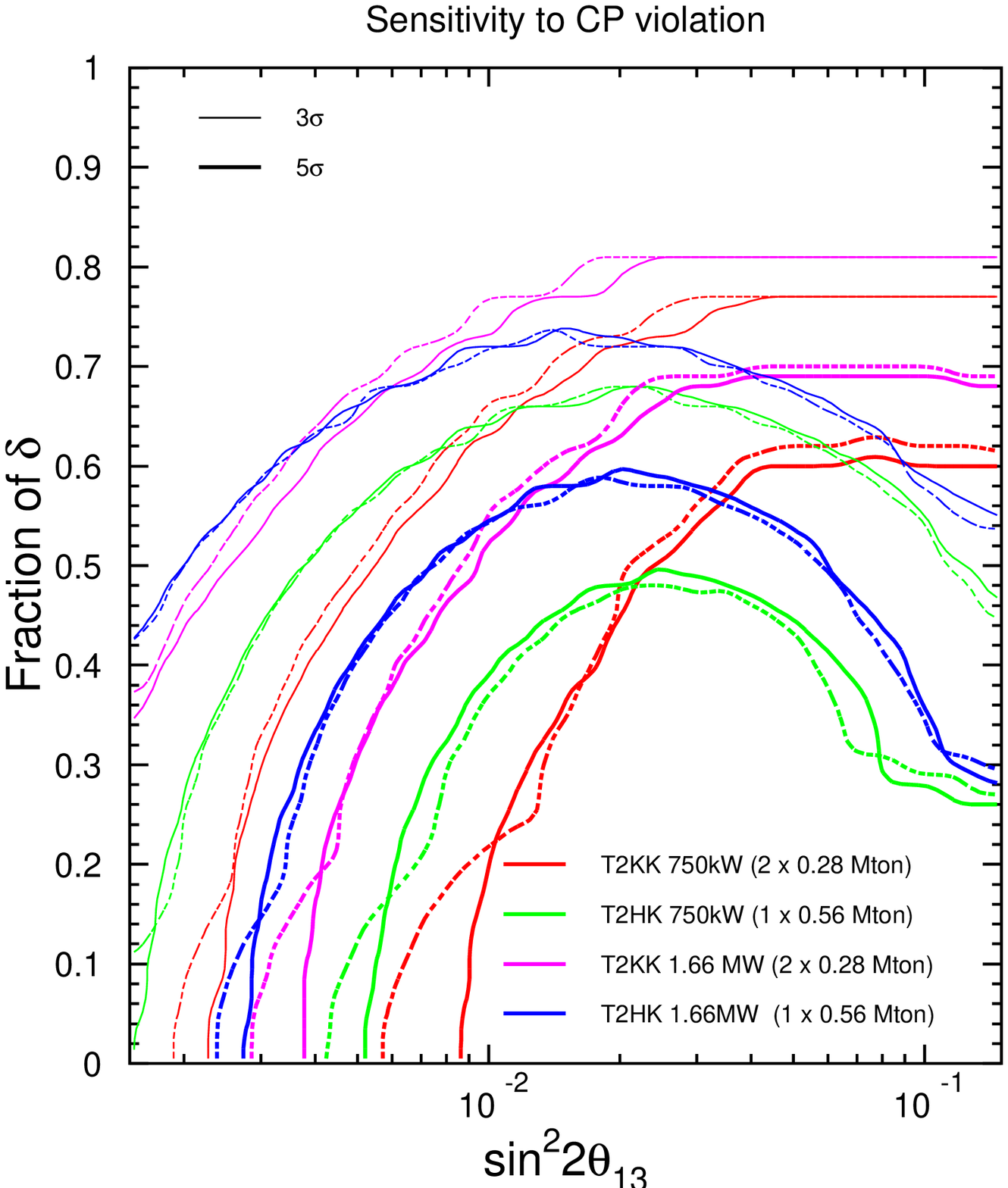}
}} \vspace{-1.5pc} \caption{{\bf Left:} Region of CP phase and
  $\theta_{13}$ where CP violation can be found. {\bf Right:} Fraction
  of $\delta_{CP}$ phase space in for which CP violation can be
  found. (Plain lines = normal hierarchy, dashed
  lines = inverted hierarchy)}
\label{fig:cp}
\end{figure*}

It was shown in the Hyper-Kamiokande Letter of Intent that the best
results are obtain when running with three years of neutrinos and
seven years of anti-neutrinos so I kept this running ratio for the
Hyper-Kamiokande simulation. For the T2KK simulation I used five years
of neutrinos and five years of anti-neutrinos as it was shown to be
the best previously~\cite{Ishitsuka:2005qi}. Figures~\ref{fig:mass}
and Figures~\ref{fig:cp} show the capability of T2KK versus T2HK for a
750~kW beam and a 1.66~MW beam. Note that I assumed an unknown mass
hierarchy when measuring the CP phase. In Figures~\ref{fig:mass} we
can see that T2KK is able to solve the mass hierarchy at $5 \sigma$
regardless of the value of $\delta_{CP}$ and even for a 750~kW
beam. In Figures~\ref{fig:cp}, we see that the T2KK setup can also
determine whether CP violation exists for a larger fraction of the
$\delta_{CP}$ phase space than T2HK. T2KK covers 60\% of the phase
space at $5 \sigma $ for a 750~kw beam and 70\% for a 1.66~MW
beam. However let me remind that in this study it was assumed that the
mass hierarchy is not known even when trying to measure the CP
phase. If the mass hierarchy is known the results for T2HK are much
better as can be seen in the Hyper-Kamiokande Letter Of
Intent~\cite{Abe:2011ts}.

\section{Conclusions}

Given the large value of $\theta_{13}$ measured in 2012, the Tokai to
Kamioka and Korea (T2KK) setup with a 750~kW beam is able to solve the
mass hierarchy at $\sigma$, for any value of the CP phase. It also
tells us if there is CP violation in the lepton sector for 60\% of the
values of the CP phase.  It is important to note that as long as the
mass hierarchy is unknown, T2HK alone cannot achieve this. This is
because the longer baseline to Korea is needed to solve the mass
hierarchy and solve the degeneracies of the parameter space. However
if the mass hierarchy is known then T2HK is equivalent to T2KK for the
measurement of the CP phase.

\bibliographystyle{aipproc}   

\bibliography{t2kk12_biblio}

\IfFileExists{\jobname.bbl}{}
 {\typeout{}
  \typeout{******************************************}
  \typeout{** Please run "bibtex \jobname" to optain}
  \typeout{** the bibliography and then re-run LaTeX}
  \typeout{** twice to fix the references!}
  \typeout{******************************************}
  \typeout{}
 }

\end{document}